# Electromagnetic acceleration of permanent magnets

*S. N. Dolya*

*Joint Institute for Nuclear Research, Joliot - Curie 6, Dubna, Russia, 141980*

**Abstract**

We consider the acceleration of the permanent magnets, consisting of NdFeB by means of the running magnetic field gradient. It is shown that the specific magnetic moment per nucleon in NdFeB is determined by the remained magnetization of the substance and for NdFeB it is equal to the $m_{NdFeB} = 1.57 * 10^{-10}$ eV / Gs. The maximum accessable gradient of the magnetic field accelerating the permanent magnets is determined by the coercive force $H_c \sim 30$ kGs. For the NdFeB magnets this gradient is equal to: $dB_z / dz = 20$ kGs / cm. The finite velocity of the magnets $V_{fin} = 6$ km / s, the length of acceleration is $L_{acc} = 637$ m.

**1. Introduction**

The acceleration of the magnetic dipoles can be carried out in the Gauss gun [1], wherein the iron core is magnetized by the field coil with a current and further it is involved into the coil. When passing of the iron core through the coil centre, the current in the coil abruptly stops and the iron core is flying to the next coil where the same process takes place.

The Gauss gun has a number of drawbacks. First of all, while driving of the core through the Gauss gun the core will deviate exponentially from the selected synchronous phase since the phase motion in the Gauss gun in unstable [2]. This means that if we calculate the movement of the core for some position of the core relatively the position of the current pulse in the coil, which is called the phase, then we find that the core while its moving along the coils with the current deviates exponentially from the phase of the current in the coil.

Indeed, let the core be a little bit behind the phase in its movement and it has arrived later to the coil where the current had already shut down. This means that the core attains less energy during its passage through this coil and its being late will grow with every following coil.

On the contrary, if the core comes ahead to some coil, it gets more energy and to the following coil it will come with a growing advance.

Physically this means that the region of stable phases is located in the front slope of the pulse, and steadily to accelerate the body it is possible only by pushing the body with the current pulse. It is not possible steadily to accelerate the core if the pulse pulls the core [3]. This is easier to understand trying to speed up the iron core by means of a little magnet. This little magnet should be taken into the hand and it will imitate the current pulse running on the coils. The iron core is magnetized and attracted to the little magnet in the hand.

The hand with this magnet must move with acceleration imitating the current pulse running through the coils of the Gauss gun. Then it will be clear that it is impossible to reach the stable acceleration. If the hand moves too fast, the iron core will get into the less and less field gradient of the little magnet and the core will be behind the hand. If you start to move the hand a little bit slower, the iron core, approaching the hand with the little magnet will get into the increasing gradient of the magnetic field. It will accelerate faster and, finally, will catch up with the hand holding the little magnet. Thus, it is not possible to obtain the stable acceleration in this case.

The situation is different if the hand with a little magnet pushes the other little magnet. When approaching each other the repulsive force between these little magnets increases and the little magnet kept in the hand faster and faster accelerates the other little magnet. If the little magnet goes too far from the one in the hand it will get into the region of the less gradients of the magnetic field. The little magnet in the hand is moving with acceleration and that is why it will catch up with the other little magnet which is accelerated. In this case we will observe the process of sustainable acceleration.

As we have mentioned above, this acceleration picture is explained by the following: the region of stable phases of acceleration is located on the front slope of the accelerating current pulse and it is necessary to push the little magnet being accelerated but not pull it.

Another disadvantage of the Gauss gun is the low rate of acceleration of the magnetic dipole in it.

The force of the magnetic dipole interaction $F_z$ with the gradient of the magnetic field can be written as follows:

$$F_z = m*dB_z/dz, \qquad (1)$$

where m - the magnetic moment per mass unit, $dB_z/dz$ - the magnetic field gradient.

The accelerator technology chooses the nucleon mass as the mass unit which is equal to $mc^2 \approx 1$ GeV in energy units.

In iron the magnetic moment per atom is [4] p. 524, m = 2.22 $n_b$, where $n_b = 9.27 * 10^{-21}$ erg / Gs, the Bohr magneton, [4], p. 31. Taking into account that the atomic weight of iron A is approximately equal to A = 56 and going from erg units to the eV units (1 erg = $6.24 * 10^{11}$ eV) it can be found that the magnetic moment per nucleon in iron is equal to the following:

$$m_{Fe} = 2.3*10^{-10} \text{ eV/Gs*nucleon.}$$

We assume that it is possible to create a running gradient of the magnetic field equal to the value of $dB_z / dz = 20$ kGs / cm = $2 * 10^6$ Gs / m. The finite velocity of the core $V_{fin} = 6$ km / s, expressed in the units of the velocity of light in vacuum, will be equal to: $\beta_{fin} = V_{fin} / c = 2 * 10^{-5}$, where c = $3*10^{10}$ cm / s – is the velocity of light in vacuum. The finite energy per nucleon in the iron core will be equal to:

$$W_{fin} = mc^2 \beta^2_{fin}/2 = 10^9*4*10^{-10}/2 = 0.2 \text{ eV/nucleon.}$$

The acceleration rate is $\Delta W / \Delta z = F_z = m * dB_z / dz = 2.3 * 10^{-10} * 2*10^6 =$
$= 4.6 * 10^{-4}$ eV / m * nucleon and the core will reach the finite energy $W_{fin} = 0.2$ eV / nucleon at the length of the accelerator equal to the following:

$$L_{acc\ Fe} = W_{fin}/(\Delta W/\Delta z) = 0.2/4.6*10^{-4} = 435 \text{ m.}$$

Note that it is not possible to sustainably accelerate the iron core in the Gauss gun.

To increase the magnetic moment per nucleon in comparison with the iron is possible if to use the superconducting coil with the current as the body being accelerated [3]. For the coil with a diameter of $d_{turn} = 6$ cm it is possible to increase the magnetic moment per nucleon by 17 times.

**2. The interaction of the running magnetic field gradient with a constant magnet based on NdFeB**

It is needed to preserve the superconductivity inthe current coil while accelerating as well as during the time of flight of the accelerated magnetic dipole to the target. It is not convenient. We consider a possibility of accelerating the permanent magnet. The most powerful permanent magnet is based on the composition of the magnets made of NdFeB.

We assume that the magnets are made of 30% of neodymium with atomic mass $A_{Nd} = 144$ and 70% of the iron having the atomic mass $A_{Fe} = 56$. Find the average atomic mass $A_{NdFeB} = 144 * 0.3 + 0.7 * 56 = 82$. The density of the composition of NdFeB is taken to be equal $\rho_{NdFeB} = 7.4$ g / cm$^3$. Then, from the relation of Avogadro we find the number of atoms per cubic centimeter:

$$6*10^{23} \ldots\ldots\ldots\ldots 82 \text{ g}$$
$$x \ldots\ldots\ldots\ldots 7.4 \text{ g}.$$

From the above we find: $n_{NdFeB} = 5 * 10^{22}$ atoms in a cubic centimeter. The role of the magnetic field of saturation will now be played by $B_r$ - the residual magnetization of the sample consisting of NdFeB. Dividing the residual induction by $4\pi$ and the density of atoms $n_{NdFeB} = 5 * 10^{22}$, we get the magnetic moment per atom of the substance. Let us take a residual induction $B_r$ to be equal to $B_r = 13.5$ kGs.

Then we find that the magnetic moment per atom in the substance NdFeB is roughly the same as in iron:

$$n_{NdFeB} = 1.35*10^4/(12.56*5*10^{22}) = 2.21 n_b.$$

However, due to the fact that the atomic weight NdFeB is greater than that of the iron and, therefore, it contains a larger number of nucleons, the magnetic dipole moment per nucleon in NdFeB, turns out to be smaller than that of the iron and is equal to:

$$m_{NdFeB} = 1.57*10^{-10} \text{ eV/Gs}.$$

This means that the rate of accumulation of energy in the running gradient of the magnetic field $dB_z / dz = 20$ kGs / cm in this case will be equal to:
$\Delta W / \Delta z = F_z = m * dB_z / dz = 1.57 * 10^{-10} * 2 * 10^6 =$
$= 3.14 * 10^{-4}$ eV / (m * nucleon). The finite energy $W_{fin} = 0.2$ eV / nucleon will be obtained by the NdFeB permanent magnet at the following length of acceleration:

$$L_{acc} = W_{fin}/(\Delta W/\Delta z) = 0.2/3.14*10^{-4} = 637 \text{ m}.$$

**3. Acceleration of the permanent magnets in a spiral waveguide**

The Gauss gun, in fact, is an artificial line with definite parameters, i.e., in this line there are individual capacitors which discharge onto separate inductors. If you transfer from the line with the definite parameters to the line with distributed parameters which is a spiral waveguide, it is possible to significantly increase the average rate of the energy gain by the permanent magnet. This will be obtained due to the fact that the gaps between the coils where the magnet moves without acceleration will be excluded.

At the same time, as usually for linear particle accelerators we have to choose a synchronous phase on the front slope of the current pulse and carry out a detailed calculation of the dynamics of the magnet being accelerated. This calculation for the magnet containing an iron core and a superconducting coil was performed in [5].

Below we consider the acceleration of the permanent magnet made of NdFeB, having a length of $l_{NdFeB} = 1$ cm. We assume that the acceleration of the permanent magnet is performed by the running gradient of the magnetic field. This gradient is generated by the running current pulse via the coils of the spiral.

As in [5] we obtain the relation between the value of the gradient and the value of the magnetic field - $dB_z / dz \ B_z$:

$$dB_z/dz = k_3 B_z = (2\pi/\lambda_s) B_z, \qquad (2)$$

where $k_3$ (the wave vector in the z-direction) is equal to: $k_3 = 2\pi / \lambda_s$, $\lambda_s$ – the slowed down wave length propagating in the spiral waveguide is $\lambda_s = \lambda_0 * \beta_{ph}$; $\lambda_0 = c / f$ - the wavelength in vacuum; $c = 3*10^{10}$ cm / s is the velocity of light in vacuum; f is the frequency of the wave. The wave frequency f is related to the duration of the current pulse propagating along the coils of the spiral with $\tau$ ratio: $f \approx 1 / 2\tau$.

The phase velocity of the wave $\beta_{ph}$, expressed in units of the velocity of light, must coincide with the velocity of the accelerated permanent magnet. At the end of acceleration this velocity must be equal to $\beta_{ph\ fin} = \beta_{fin} = 2 * 10^{-5}$, that corresponds to the normal velocity of the permanent magnet equal $V_{fin} = \beta_{fin} * c = 6$ km / s.

Inside the spiral waveguide the components of the longitudinal electrical and magnetic fields can be written as:

$$E_z = E_0 I_0(k_1 r),$$
$$B_z = -i(k_1/k) \text{tg}\Psi I_0(k_1 r_0) E_0 I_0(k_1 r)/I_1(k_1 r_0), \quad (3)$$

where $I_0$, $I_1$ - a modified Bessel function, $r_0$ – the radius of the spiral, $k_1 = k (1/\beta^2_{ph} - 1)^{1/2}$ – the radial wave vector inside the spiral waveguide. The total wave vector $k = \omega/(c\varepsilon^{1/2})$, $\varepsilon$ – is the dielectric constant of the medium, which fills the spiral waveguide, $\omega = 2\pi f$ –is the angular frequency, $\text{tg}\Psi = h/2\pi r_0$ – is the tangent of the spiral winding angle, h - the winding pitch of the spiral. The value of $E_0$ is the amplitude of the electric field on the axis of the spiral.

For the case of $\beta_{ph} \ll 1$ it can be considered $k_1/k = 1/\beta_{ph}$. We also assume that the whole accelerator is divided into individual sections, where in each section the following condition is fulfilled: $x = k_1 r_0 = 2\pi r_0/(\beta_{ph}\lambda_0) = 1$. Bearing in mind that $I_0(1)/I_1(1) = 2$ and $I_0(0) = 1$, we get:

$$B_z/E_z = 2\text{tg}\Psi/\beta_{ph}. \quad (4)$$

Since for the dense winding of the spiral and dielectric filling of the area between the coil and the screen there is the following condition [5]:

$$\beta_{ph} = \sqrt{2}\, \text{tg}\Psi/\varepsilon^{1/2}, \quad (5)$$

we, finally, obtain:

$$B_z/E_z = (2\varepsilon)^{1/2}. \quad (6)$$

Here we must remember the following. When accelerating of the iron cores in the Gauss gun the iron core is magnetized by the current pulse accelerating this core in such a way that the core is pulled into the coil. Similarly, the core is pulled into the coil in a home electrical bell. At the acceleration of the permanent magnets by the current pulse running via the spiral coils these magnets will be pushed out from the coil by the magnetic field gradient, and they will be demagnetized by the magnetic field of the current pulse.

The iron core can be accelerated by any value of the magnetic field gradient and by the value of the magnetic field $B_{acc}$ associated with this gradient. When the induction reaches saturation the iron core simply ceases further magnetization.

If the value of the field $B_{acc}$ exceeds the coercive force of the permanent magnet $B_{acc} > H_c$, the permanent magnet is simply demagnetized. In this case the dipole magnetic moment will become zero and the permanent magnet ceases the acceleration by the running gradient of the magnetic field.

The best permanent magnets made of NdFeB have the coercive force $H_c > 30$ kGs. As it will be shown below, for the short spatial running current pulses the gradient of the magnetic field is achievable: $dB_z / dz > 20$ kGs / cm.

**4. Selection of main parameters of the spiral waveguide**

We choose the radius of the spiral waveguide winding $r_0 = 1$ cm. This means that we have chosen the slowed down wavelength equal to: $\lambda_s = 2\pi r_0 = 6$ cm and this wavelength will remain unchanged throughout the accelerator. Take the initial velocity of acceleration of the permanent magnets to be equal to the following: $V_{in} = 600$ m / s, [5].

This velocity is assumed to achieve by the preliminary gas-dynamic acceleration of permanent magnets. Thus, the electromagnetic acceleration is performed from the initial velocity, expressed in terms of the velocity of light $\beta_{in} = 2 * 10^{-6}$ to the final velocity $\beta_{fin} = 2 * 10^{-5}$.

We find the spatial length of the current pulse accelerating the permanent magnets. The wavelength $\lambda_0 = \lambda_s / \beta_{ph}$ and it is equal to $\lambda_{0\ in} = 3 * 10^6$ cm to start acceleration and $\lambda_{0\ fin} = 3 * 10^5$ cm - for the end of acceleration. To support the optimal condition between the power involved into the spiral waveguide and the amplitude $E_0$ of the electrical field on the acceleration axis [6], it is necessary to alter the length of the accelerating current pulse $\tau$ from section to section in the process of acceleration.

For the start of acceleration the wave length $\lambda_{0\ in} = c * 2\tau_{in}$, from where $\tau_{in} = \lambda_{0\ in} / 2c = 50$ μs. At the end of acceleration the duration of the current pulse accelerating the permanent magnet must be 10 times shorter: $\tau_{fin} = \lambda_{0\ fin} / 2c = 5$ μs. This pulse duration corresponds to the frequency of the acceleration at the start of acceleration $f_{in} = 1 / 2\tau_{in} = 10^4$ Hz, and at the end of acceleration it corresponds to $f_{fin} = 1 / 2\tau_{fin} = 10^5$ Hz.

From (2) we find the magnetic field value corresponding to the magnetic field gradient $dB_z / dz = 20$ kGs / cm. For $\lambda_s / 2\pi = 1$ cm the magnetic field of the running pulse is: $B_z = 20$ kGs.

Knowing the magnetic field of the pulse from the formula (6) we can find the electric field of the pulse on the system axis $E_z = B_z / (2\varepsilon)^{1/2}$. Choose the value of $\varepsilon$ for the medium, which fills the space between the coil and the outer screen $\varepsilon = 1.28 * 10^3$. This value of $\varepsilon$ is contained, for example, in barium titanate. From the electro-technical point of view this type of filling the space between the coil and the outer screen means that we have increased the capacity between the coils of the spiral and the capacitance between the coils of the spiral and the screen by $\varepsilon$ times.

Introduction of the medium with a high value of $\varepsilon$ allows one additionally to slow down the wave accelerating the permanent magnets.

Substituting the numbers into the formula (6) we find that $E_z = 120$ kV / cm. This value is smaller than the amplitude value $E_z$ of the electric field on the axis $E_0$. It is smaller because we have to use a non-maximum value of the electrical field. [2]. Now we choose $\sin\varphi_s = 0.7$, that means that we have chosen the synchronous phase $\varphi_s = 45^0$.

The amplitude value of the field $E_0$ in this case turns out to be equal to:

$$E_0 = E_z/\sin\varphi_s = 170 \text{ kV/cm}.$$

We can use the following formula to estimate the value of the pulse voltage which will run in the spiral and create an appropriate electric field $E_0$:

$$U = E_0 * \lambda_s/2\pi = 170 \text{ kV}.$$

From the relation between the power of the energy flow and the field strength on the axis of the spiral wave guide we can find the required pulse power [5]. It is equal to the following:

$$P = (c/8)E_0^2 r_0^2 \beta_{ph}\{(1+I_0K_1/I_1K_0)(I_1^2 - I_0I_2) + \varepsilon(I_0/K_0)^2(1+I_1K_0/I_0K_1)(K_0K_2 - K_1^2)\}. \qquad (7)$$

The argument of the modified Bessel functions of the first and second kind shown in the curly brackets is the value of $x = 2\pi r_0 / \lambda_s$, which we have chosen to be equal to: x = 1. Then, for this argument, the second term in brackets is much larger than the first one, and the bracket itself is: { } = 3.77 * ε, [6]. Substituting the numbers into the formula (7), we find:

$$P (W) = 3*10^{10}*1.7*1.7*10^{10}*2*10^{-5}*3.77*1.28*10^{3}/(8*300*300*10^{7}) =$$
$$=12.3 \text{ MW}.$$

The issues of the pulse attenuation while its propagating in the spiral waveguide will not be considered in detail. We will not discuss the following issues: the capture of permanent magnets into the acceleration mode, the accuracy of the synchronization of start of the gas-dynamic acceleration and of the phase of the current pulse propagating along the spiral. All these issues as well as requirements to the accuracy of maintaining the parameters of acceleration, release of the permanent magnets into the atmosphere of the Earth, etc, are discussed it in detail in [5-8].

Let us concentrate only on the issues of focusing the permanent magnets, i.e., how to hold the permanent magnet near the axis of acceleration. You can continue the analogy with the acceleration of the little permanent magnet by using a little magnet in the hand. You can see that at the repulsion of the little magnet being accelerated from the magnet clutched in hand, the little accelerated magnet tends to deviate from the axis of the acceleration in the transverse direction. It also tends to turn by180 degrees to be pulled to the little magnet clutched in the hand.

Deviation from the acceleration axis is explained by the fact that in the azimuthally symmetric acceleration field it is impossible to fulfill simultaneously the conditions of the radial and phase stability [2]. The radial instability corresponds to the area of phase stability, which we have chosen because we are pushing the permanent magnet. This means that the accelerating field pushes out the magnet being accelerated from the acceleration axis. Note that in the Gauss gun where the running current pulls the core, this current coil pulls the core into the center, i.e., it puts it onto the acceleration axis.

For the magnetic and electric dipole there is the force which turns the dipole by 180 degrees. There is no such force while accelerating the charged particles.

To keep the magnetic dipoles near the axis of the acceleration, it is necessary to introduce additional external fields or establish the conditions under which the dipoles will be kept near the axis of the acceleration.

In [5] this holding of the magnetic dipoles near the axis of the acceleration is performed by means of using the magnetic field which alters its sign in the space. In [3, 8] it is proposed to introduce acceleration in the narrow channel to keep the accelerated body near the acceleration axis. There is no opportunity for the body under acceleration to turn in this narrow channel by 180 degrees and this channel will also hold the body near the axis of acceleration.

## 5. Conclusion

When we develop the permanent magnets with residual induction greater than, $B_r > 13.5$ kGs it will be possible to achieve a more specific magnetic moment per nucleon than in the NdFeB magnets:
$m > m_{NdFeB} = 1.57 * 10^{-10}$ eV / Gs. If you can get a great coercive force of the magnets than $H_c = 30$ kGs, it will be possible to increase the magnetic field gradient accelerating the magnetic dipoles by more than 20 kGs/cm.

http://arxiv.org/ftp/arxiv/papers/1312/1312.7393.pdf